\documentclass{PoS}
\newcommand{\met}{\rlap{\kern0.25em/}E_T}

\title{Higgs Searches at the Fermilab Tevatron $p\bar{p}$ Collider}

\ShortTitle{Higgs Searches at the Tevatron}

\author{\speaker{Jianming Qian}\thanks{On behalf of the CDF and D\O\ Collaborations}\\
        Department of Physics, University of Michigan, Ann Arbor, Michigan 48109, USA\\
        E-mail: \email{qianj@umich.edu}}

\abstract{Searching for and potential discovery of Higgs boson(s) both within and beyond the standard model is perhaps the most visible physics goal of the current Fermilab Tevatron program. In this proceeding, recent results from both the CDF and D\O\ experiments based on analyses of datasets with integrated luminosities between 1.7 and 3.0 fb$^{-1}$ are summarized. The combined Tevatron cross section upper limits on the production of a standard model Higgs boson are fast approaching the expected standard model values for a wide mass range. Particularly, the Tevatron has now excluded a standard model Higgs boson with a mass of 170~GeV at 95\% C.L.}

\FullConference{2008 Physics at LHC\\
		 September 29 - 4 October 2008\\
		 Split, Croatia}

\begin{document}

\section{Introduction}
The standard model of particle physics has been remarkably successful in interpreting existing experimental data. However, the electroweak symmetry breaking of the model remains to be understood. In the model, the symmetry is postulated to be spontaneously broken through the so-called Higgs mechanism~\cite{higgs}. As a result, a neutral scalar fundamental particle called the Higgs boson is predicted to exist. Searching for this elusive particle has been a major undertaking of the experimental particle physics community over the last two decades. 

With the exception of its mass, the couplings of the Higgs boson to fermions and gauge bosons are theoretically well known. Its mass, however, can be inferred from precision measurements of electroweak data, as the Higgs boson often appears in the loops of high-order electroweak corrections.  A recent global fit to precision electroweak data yields a central value for the Higgs boson mass of $84^{+34}_{-26}$~GeV and a 95\%~C.L. upper bound of 154~GeV~\cite{ewwg}. Direct searches at LEP have set a lower bound of 114.4~GeV on the mass, again at 95\% C.L.~\cite{leplimit}.

The relatively low mass preferred by the fit is encouraging for the search for and potential discovery of the Higgs boson at the Fermilab Tevatron $p\bar{p}$ collider. At the Tevatron, the Higgs boson can be produced through the following four processes (ordered in decreasing cross section): (a) $gg\to H$, (b) $q\bar{q}'\to WH$, (c) $q\bar{q}\to ZH$ and (d) $qq\to qqH$. Since it couples to other particles in proportion to the mass of the particle, the Higgs boson decays predominantly to the heaviest particle kinematically allowed. Thus in the low mass region of $114<m_H\stackrel{<}{_\sim}135$~GeV, Higgs decays mostly to $b\bar{b}$, while for masses above 135 GeV, $H\to WW^*$ decay dominates. For example, $Br(H\to b\bar{b}) = 73\%$ at 115~GeV and $Br(H\to WW^*) = 97\%$ at 170~GeV.

The minimal extension of the standard model, the minimal supersymmetric standard model (MSSM), predicts five Higgs bosons: two CP even ($h$, $H$) and one CP odd ($A$) neutral bosons as well as two charged bosons ($H^+$ and $H^-$)~\cite{mssm}. Two of the three neutral bosons are generally degenerate in mass and are often collectively denoted as $\phi$. Two main production processes for $\phi$ at the Tevatron are $b\bar{b}\to \phi$ and $gg\to\phi$. Their cross sections are proportional to $\tan^2\beta$, here $\tan\beta$ is the ratio of the vacuum expectation values of the two Higgs doublet fields in the model. Thus at high $\tan\beta$, the cross section could be large. Moreover, $\phi$ decays mostly to $b\bar{b}$ and $\tau\tau$. At $\tan\beta \sim 40\ (\approx m_t/m_b)$, for example, $Br(\phi\to b\bar{b}) \approx 90\%$ and $Br(\phi\to \tau\tau)\approx 10\%$.

The results summarized in this proceeding are from the CDF and D\O\ experiments at the Fermilab Tevatron $p\bar{p}$ collider at $\sqrt{s}=1.96$ TeV. The Tevatron has been running well with over 5 fb$^{-1}$ integrated luminosity delivered to each experiment. Analyzing these huge datasets takes time. Consequently the results described below are from data samples with luminosities between $1.7-3.0$~fb$^{-1}$. The CDF and D\O\ detectors are designed to explore high $p_T$ phenomena and are described in detail elsewhere~\cite{det}. 

The expected low rate of a potential Higgs signal poses a number of experimental challenges including, but not limited to, event selection efficiency, $b-$tagging performance and dijet mass resolution. Both the CDF and D\O\ experiments have invested considerable effort in all these areas. For example, CDF utilizes tracking and calorimetry information to significantly extend their muon identification capability, and uses their central tracking to supplement the calorimetry to improve the dijet mass resolution.  D\O\ employs a neural network to combine several $b-$tagging algorithms to increase the $b-$tagging efficiency by over 30\% for a fixed fake rate.   
The maturing understanding of the datasets enables both collaborations to utilize advanced analysis techniques such as a neural network (NN), matrix-element (ME), and boosted decision trees (BDT) to exploit subtle differences between the signal and background processes, and hence greatly improve search sensitivities. NN is a pattern recognition paradigm that consists of a large number of interconnected processing elements trained to identify a specific pattern. The ME method on the other hand constructs likelihood discriminant using the matrix-elements (generally leading-order) of the underlying physical processes. And BDT is a classification algorithm that combines a set of weak classifiers into a powerful discriminant.

It is not possible to do justices to all the efforts that have gone into the Higgs searches in this short proceeding. Readers are encouraged to consult collaborations' web pages~\cite{webs} for a full slate of the results as well as the latest updates.

\section{Searches for a standard model Higgs boson}
Though the $gg\to H$ production dominates over the entire mass region of interest at the Tevatron, the large $H\to b\bar{b}$ decay branching ratio at low masses renders it impractical for the search in this mass region, due to an overwhelming QCD $b\bar{b}$ background. Consequently searches at low masses are generally focused on Higgs productions in association with vector bosons such as $WH$ and $ZH$. The decays of $W\to\ell\nu$ ($\ell=e,\mu$) and $Z\to\ell\ell,\ \nu\nu$ lead to distinct signatures for triggering and identification while $H\to b\bar{b}$ decay provides the benefit of the Higgs mass reconstruction. For a Higgs mass above 135 GeV, the $gg\to H$ production followed by the $H\to WW^*\to\ell\ell^\prime\nu\nu$ decay offers the most promise for a potential Higgs discovery, as backgrounds are strongly suppressed by the presence of two high $p_T$ leptons and a large missing transverse energy ($\met$).

\subsection{$WH\to\ell\nu b\bar{b}$}
\begin{figure}[h]
\centerline{\includegraphics[width=0.40\textwidth]{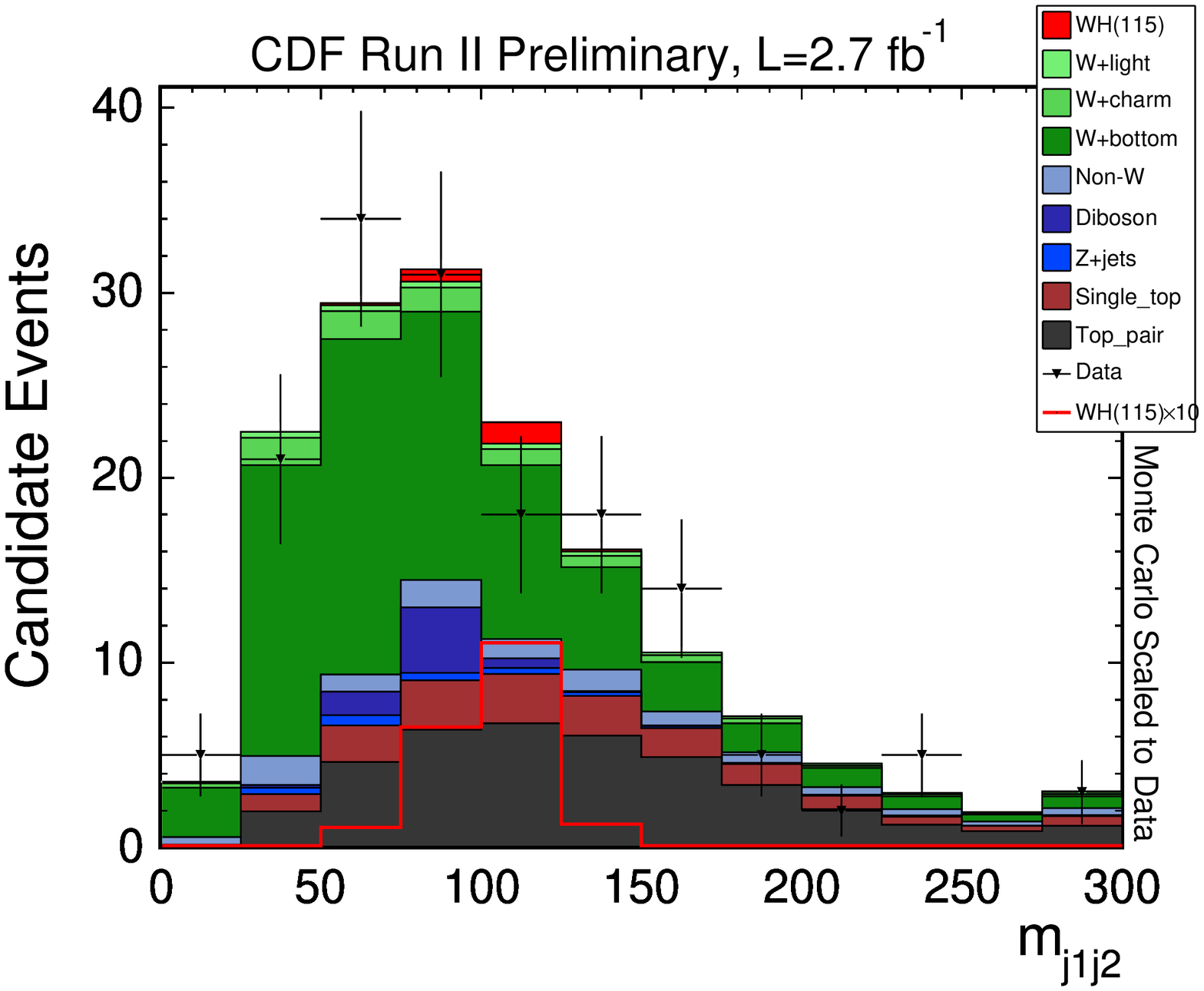}\hspace{1.5cm}
            \includegraphics[width=0.40\textwidth]{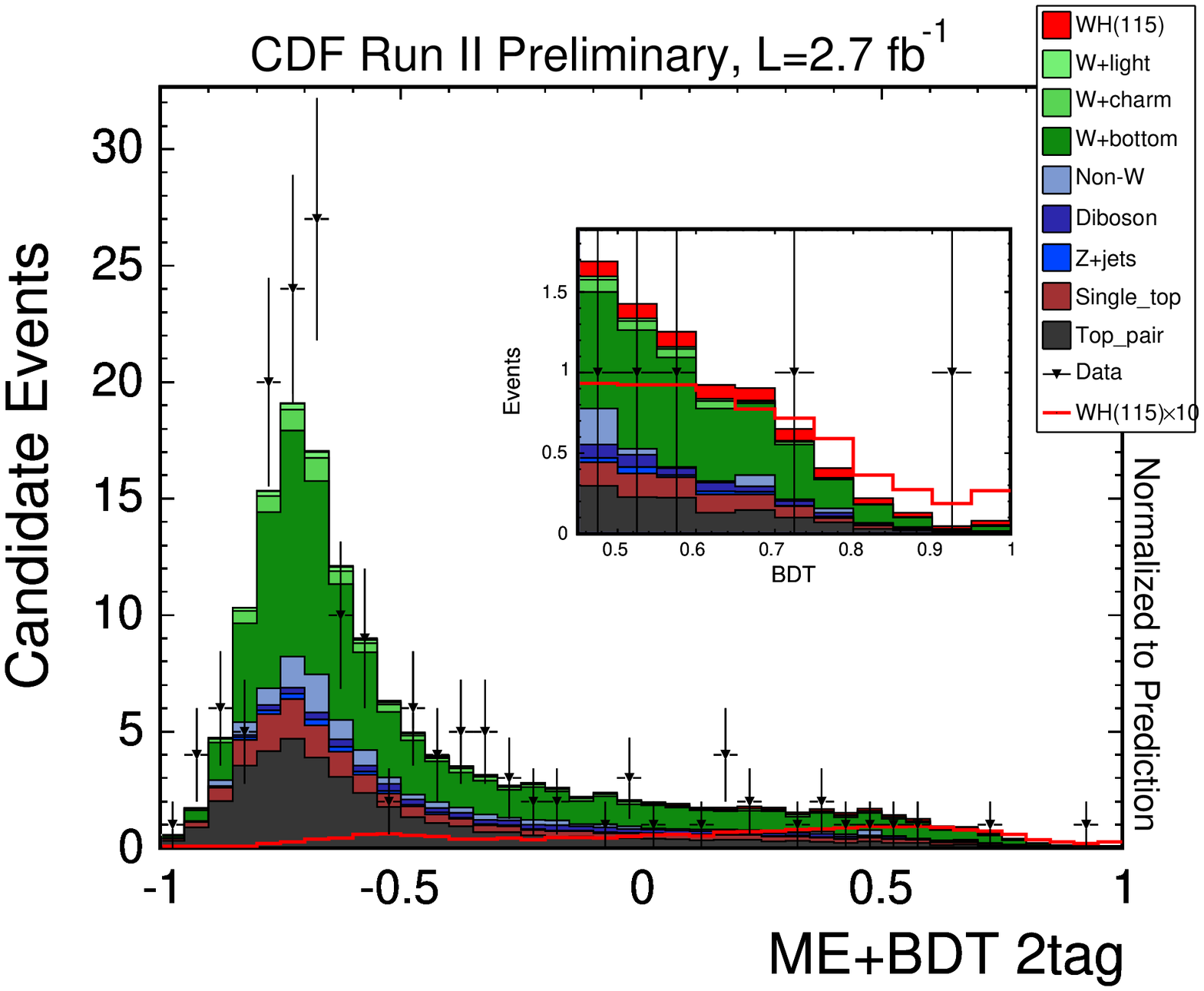}}
\caption{The invariant mass of the dijet (left) and the corresponding BDT outputs (right) of events with two $b-$tagged jets from CDF. For $m_H=115$ GeV, $2.0\pm 0.2$ Higgs events are expected in this sample.}
\label{fig:wh}
\end{figure}

The $q\bar{q}^\prime\to WH \to \ell\nu b\bar{b}$ final state is characterized by one high $p_T$ lepton, a large $\met$ and a pair of $b-$jets. At 115 GeV, the expected cross section is about 30~fb. Main background processes include $W+$jets ($Wbb$, $Wcc$, $Wjj$) , $t\bar{t}$, diboson ($WW$, $WZ$, $ZZ$) and multijet productions. The analyses generally begin with a pre-tagged sample of events with a $W$ and two or more jets. The large statistics of this sample allows for the normalizations of different contributions. Since the $W+$jets cross section is not well known, its overall normalization is fixed to the data, after subtracting contributions from other processes such as $t\bar{t}$, diboson, multijets etc. The relative contributions of $Wbb$, $Wcc$ and $Wjj$ of the $W+$jets component are determined from Monte Carlo~(MC). The multijet background is determined from the data. $b-$tagging algorithms are then applied to the data to yield samples with 1-tagged and 2-tagged events. MC background processes are subjected to tag-rate-functions of $b-$tagging algorithms to determine their contributions in these tagged samples. Advanced techniques are then employed to further separate the signal from backgrounds. After $b-$tagging, most information about Higgs is encoded in the dijet mass distribution. Thus good dijet mass resolution is critical for this analysis. Figure~\ref{fig:wh}~(left) shows the dijet mass distribution of the 2-tagged sample from CDF. There is no apparent excess in the distribution. The corresponding BDT output including ME as an input is shown in Fig.~\ref{fig:wh}~(right). Higgs candidates are expected to populate to the right of the distribution. An upper limit on the signal cross section can then be extracted by fitting the observed BDT distribution in the data to the sum of distributions expected from background processes. CDF has analyzed a dataset of 2.7 fb$^{-1}$ with both NN and BDT as the final discriminants and observed an upper cross section limit of 5.0 times the expected standard model value for $m_H=115$~GeV. D\O, on the other hand, has only analyzed 1.7 fb$^{-1}$ of the data using a NN discriminant. The observed upper limit is therefore considerably weaker at 9.3 times the standard model value.

\subsection{$ZH\to \nu\nu b\bar{b}$}
\begin{figure}[h]
\centerline{\includegraphics[width=0.40\textwidth]{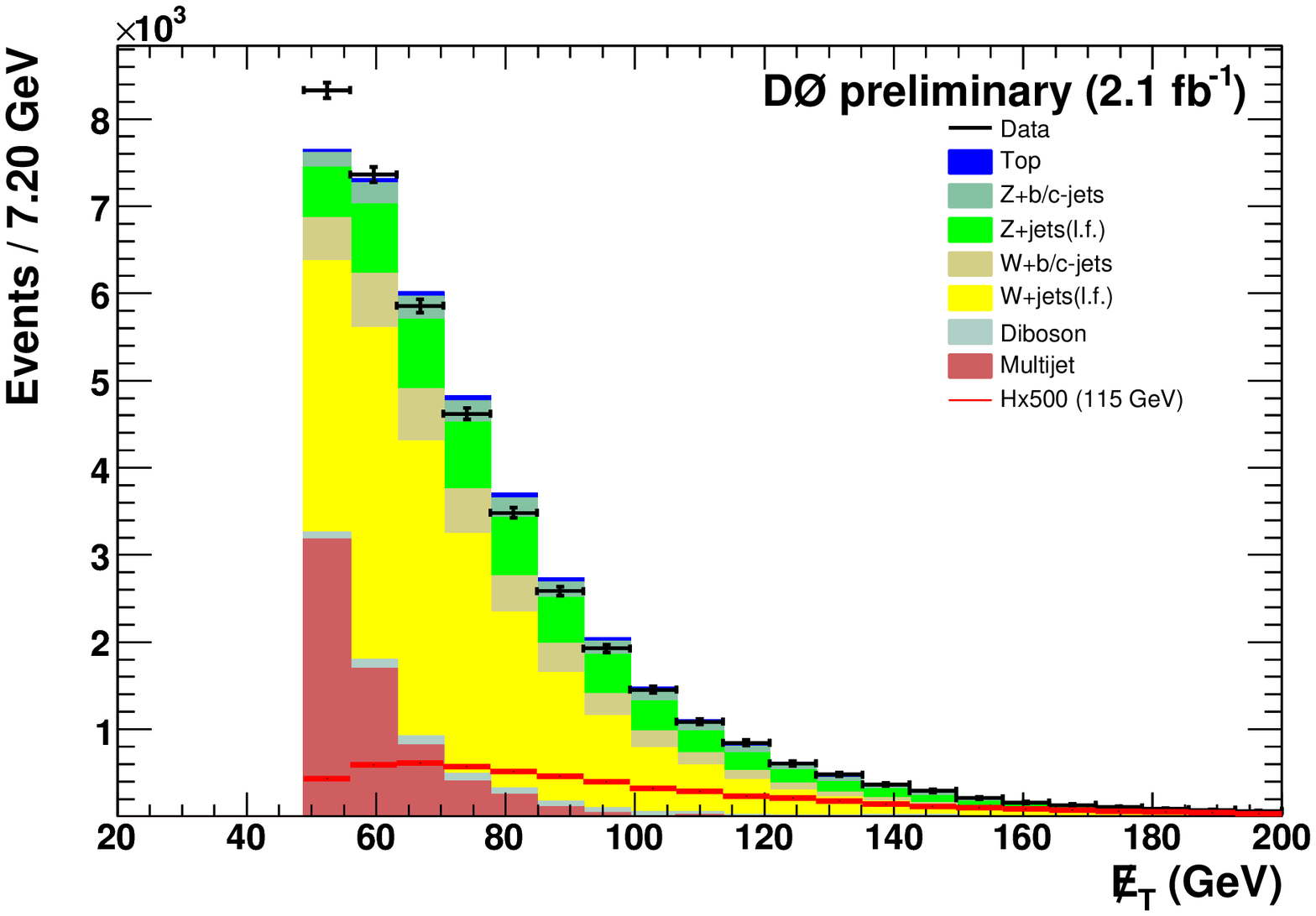} \hspace{1.5cm}
            \includegraphics[width=0.40\textwidth]{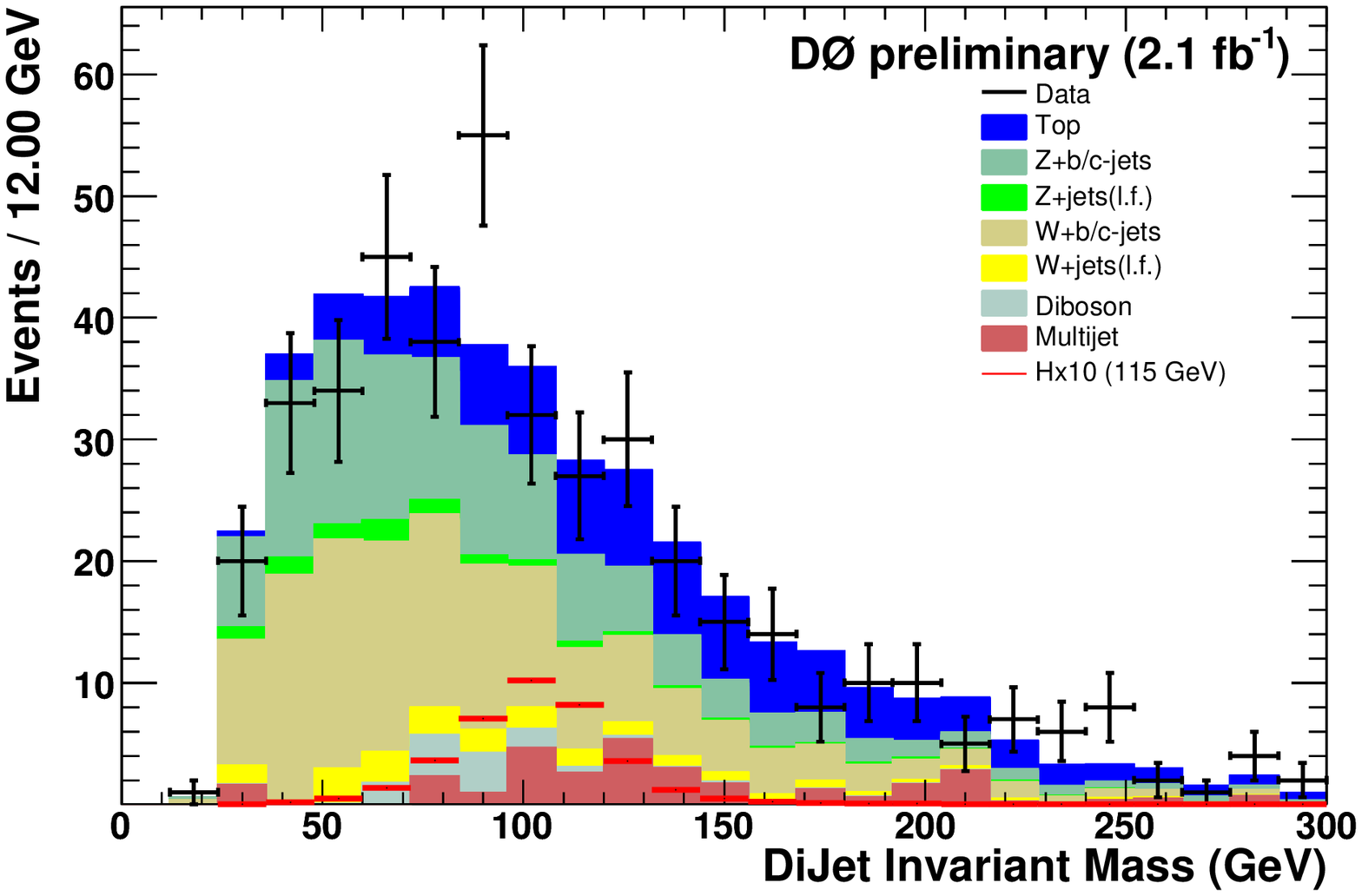}}
\caption{The $\met$ distribution before $b-$tagging (left) and the dijet mass distribution (right) for events with two $b-$tagged jets. For $m_H=115$ GeV, $\sim$ 3.7 signal events are expected in the double-tagged sample.}
\label{fig:zh}
\end{figure}

The signature of this final state is two $b-$jets and a large $\met$. The process has a cross section of $\sigma(ZH\to\nu\nu b\bar{b}) \approx 15$ fb for $m_H=115$ GeV. Before $b-$tagging, backgrounds are dominated by multijets with mis-measured $\met$ and $W+$light-jets as shown in Fig.~\ref{fig:zh}~(left) from D\O. To reduce these backgrounds, two $b-$tagged jets are required. The remaining backgrounds are $W+b/c-$jets, dibosons and $t\bar{t}$ after the $b-$tagging. Though intended initially for $ZH\to\nu\nu b\bar{b}$, the analysis is found to be also sensitive to $WH\to \ell\nu b\bar{b}$ with an identified lepton (including $\tau$). In fact, the $WH$ and $ZH$ contribute almost equally to the final signal yield. Without a high $p_T$ lepton in the final state, the $\met$ becomes the main handle for the trigger and thus a relatively high $\met$ is required. The D\O\ analysis for example requires a $\met$ to be greater than 50 GeV.  In addition, a set of topological cuts need to be applied to keep the multijet background under control. As in the case of $WH\to\ell\nu b\bar{b}$, the $ZH\to\nu\nu b\bar{b}$ production would manifest itself as a mass bump in the 2-tagged $b-$jets mass distribution.  
Figure~\ref{fig:zh}~(right) compares the observed dijet mass distribution with those from background processes, along with the expected distribution from a potential Higgs signal. Both CDF and D\O\ have analyzed 2.1~fb$^{-1}$ of the data. CDF uses a NN while D\O\ applies a BDT for the final signal-background separation. The observed upper limits for $m_H=115$~GeV are 7.9~(CDF) and 7.5~(D\O) times the expected standard model cross section.

\subsection{$ZH\to \ell\ell b\bar{b}$}
The $q\bar{q}\to ZH\to \ell\ell b\bar{b}$ process in principle provides the cleanest Higgs signature since every final-state object can be reconstructed. Unfortunately even for a Higgs process, the rate is low ($\sim 5$~fb at $m_H=115$ GeV). $Z$ boson production together with jets or another vector boson ($W$ or $Z$) constitutes the majority of the background. $t\bar{t}$ background is important as well. Candidates are selected by first identifying two leptons from $Z$ decays, followed by $b-$tagging, and finally by applying advanced discriminants. CDF has analyzed 2.4 fb$^{-1}$ of the data and used a 2-dimensional NN output to distinguish $Z+$jets and $t\bar{t}$ backgrounds from the signal. D\O\ has analyzed a dataset of similar size (2.3 fb$^{-1}$), but used a NN for the $ee$ and a BDT for the $\mu\mu$ channel as the final signal-background discriminant. At $m_H=115$ GeV, the observed upper limit in CDF is 11.6 times of the standard model value and the limit observed in D\O\ is 11.0.

\subsection{$H\to WW^*\to \ell\ell^\prime\nu\nu$}
The $gg\to H\to WW^*\to \ell\ell^\prime\nu\nu$ production and decay offers the best discovery potential at the Tevatron for a Higgs mass above 135 GeV. Compared with searches for a low mass Higgs, the direct reconstruction of Higgs mass in this final state is unfortunately infeasible due to the two undetected neutrinos. However, the final result below shows that this drawback is well compensated by the cleanliness of the event topology.  The electroweak $WW$ production, which has the identical final state as the signal, is the largest source of background. However, due to the scalar nature of the Higgs boson, $WW$ from Higgs are expected to have different kinematics. Particularly the two leptons from the Higgs decay are expected to have a smaller average opening angle in the transverse plan than those from the electroweak $WW$ process, a feature keenly exploited by both CDF and D\O. Other backgrounds include $t\bar{t}\to WWbb$ and $W+$jets with one lepton from a mis-identified jet.

\begin{figure}[h]
\centerline{\includegraphics[width=0.42\textwidth]{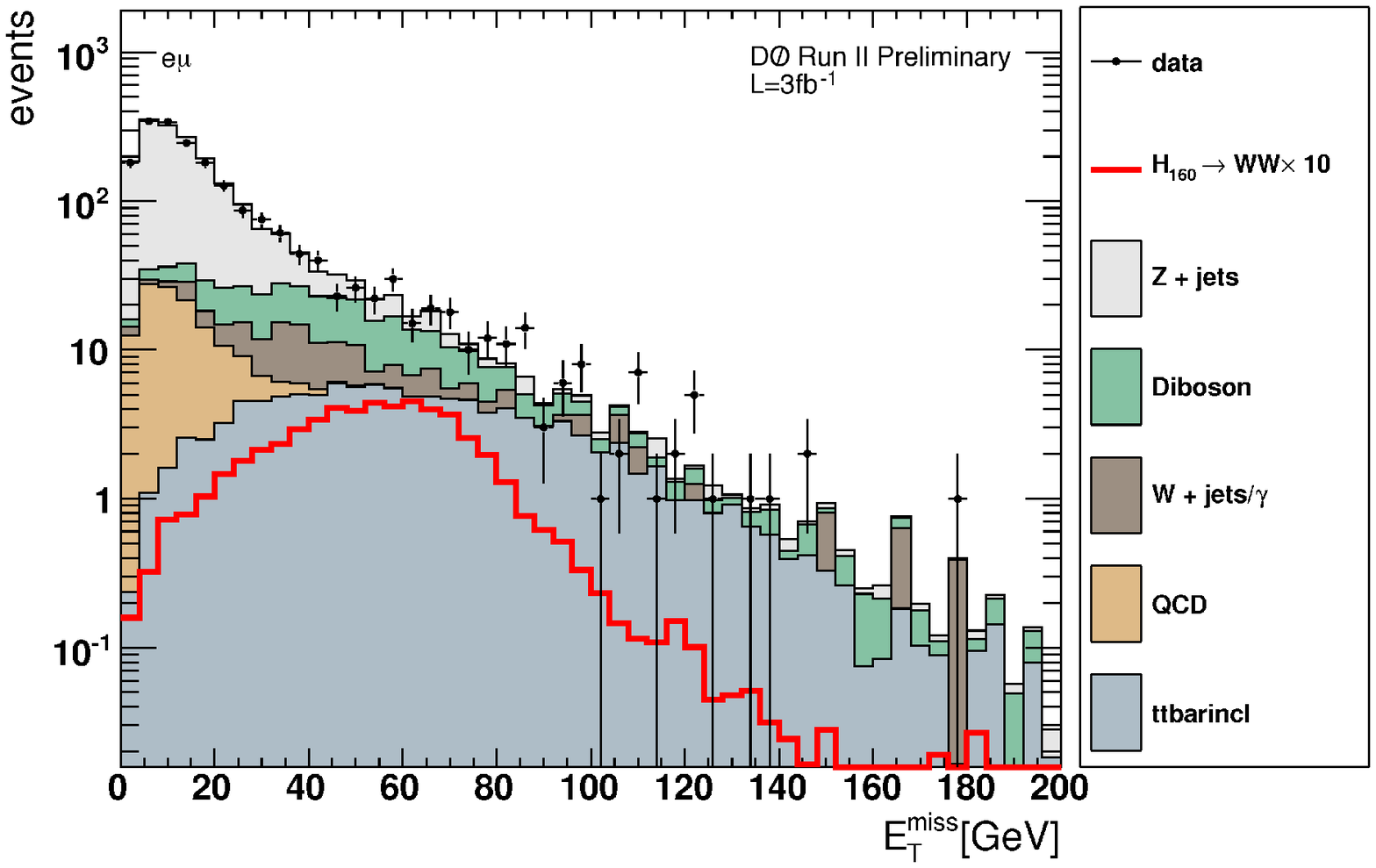}\hspace*{1.5cm}
            \includegraphics[width=0.40\textwidth]{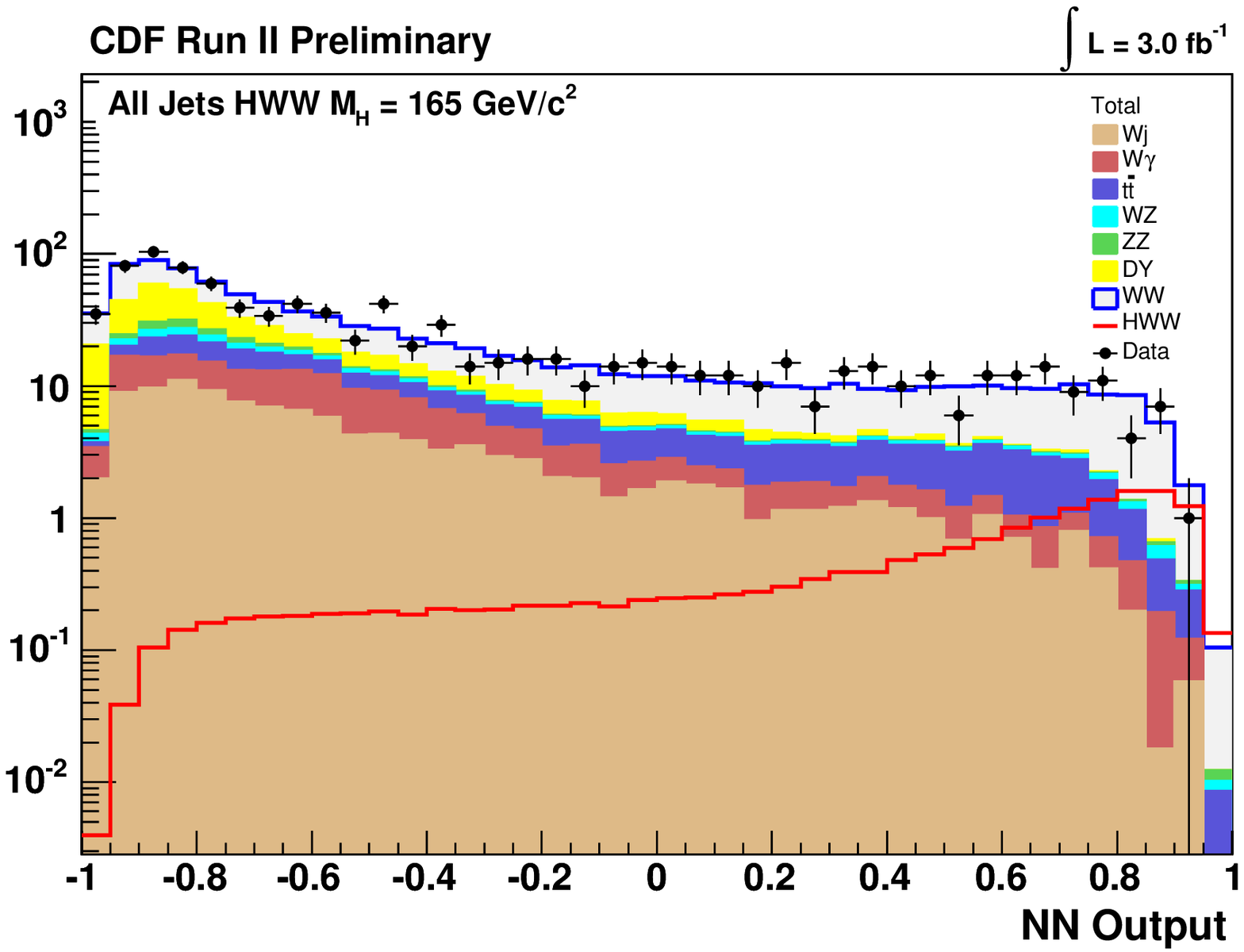}}
\caption{The $\met$ distribution of the $e\mu$ channel from D\O\ (left) and the combined NN output distribution from CDF (right).}
\label{fig:ww}
\end{figure}

Both CDF and D\O\ have analyzed 3.0 fb$^{-1}$ of the data for this final state. CDF divides their analyses into 0-, 1- and 2-jet bins, but combines $ee$, $e\mu$ and $\mu\mu$ final states. This division is motivated by the fact that signal and background kinematics are expected to be different for different jet multiplicity. Moreover, it enables the 2-jet analysis to pick up contributions from other signal processes such as $ZH\to\ell\ell b\bar{b}$ and $qq\to qqH\to qqWW^*\to qq\ell\ell^\prime\nu\nu$. Three different NNs, one for each jet bin, are trained. In contrast, the D\O\ analyses combine jet multiplicities but separate into $ee$, $e\mu$ and $\mu\mu$ channels. This approach follows naturally from the trigger and takes advantage of different signal-background ratios in the three channels. A NN is designed and trained for each channel. Efficient lepton identification and good $\met$ resolution are essential for this analysis. Figure~\ref{fig:ww}~(left) compares the observed $\met$ distribution in the $e\mu$ channel in D\O\ with expected contributions from backgrounds while Fig.~\ref{fig:ww}~(right) is the NN output distribution from CDF, combining three jet bins.  For a mass of 165~GeV, both experiments expect to see about 16 signal events. The observed upper cross section limit for the $gg\to H$ production normalized to that of the standard model value is 1.6 for CDF and 2.0 for D\O\ at $m_H=165$ GeV, by far the most stringent limits in all searches of a standard model Higgs boson.

\subsection{Other searches}
There are several other searches from both experiments that was not cover in the presentation, nor in this proceeding. The list includes $WH\to WWW^*\to\ell\ell\ell$ (CDF, D\O), $WH/ZH\to qqb\bar{b}$ (CDF), $t\bar{t}H\to t\bar{t}b\bar{b}$ (D\O), $(H\to\tau\tau)+2-$jets (CDF) and $H\to\gamma\gamma$ (D\O). Though less powerful than the searches presented above, these additional ones contribute significantly to the overall combination. Interested readers should see Ref.~\cite{webs} for details.

\subsection{Combined limits}
The limits of individual search final states presented above are all above their respective standard model cross sections. To arrive at a Tevatron limit on the Higgs production for a given mass, results from CDF and D\O\ experiments are combined, taking into account correlations in systematic uncertainties. At the time of this presentation, the two collaborations have not updated the combination for the low mass searches. However, when they are done, the combined limit is expected to be significantly stringent than the individual limits presented above.
\begin{figure}[h]
\centerline{\includegraphics[width=0.55\textwidth]{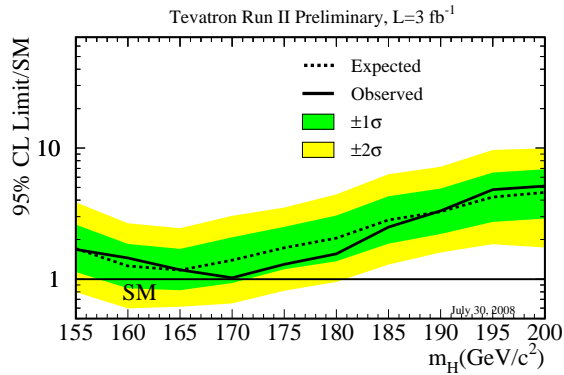}}
\caption{Observed and expected 95\% CL Tevatron combined upper limit on the Higgs cross section normalized to the standard model value as a function of the Higgs mass.}
\label{sm}
\end{figure}
For the high mass searches, the collaborations have recently combined the results of the $H\to WW^*\to\ell\ell^\prime\nu\nu$ analyses. The combined Tevatron limit is shown in Fig.~\ref{sm} as a function of the Higgs mass. It is encouraging to note that the Tevatron upper cross section limit has reached the standard model value at $m_H=170$~GeV, an indication of exciting results to come in the near future.

\section{Searches for supersymmetric Higgs bosons}
While searches for a standard model Higgs boson have the spotlight, both experiments have significant efforts in searches for the supersymmetric Higgs bosons as well. Ongoing analyses include $gb\to \phi b\to b\bar{b}b$ (CDF, D\O), $bb\to\phi\to\tau\tau$ (CDF, D\O) and $gb\to\phi b\to \tau\tau b$ (D\O).

\begin{figure}[h]
\centerline{\includegraphics[width=0.55\textwidth]{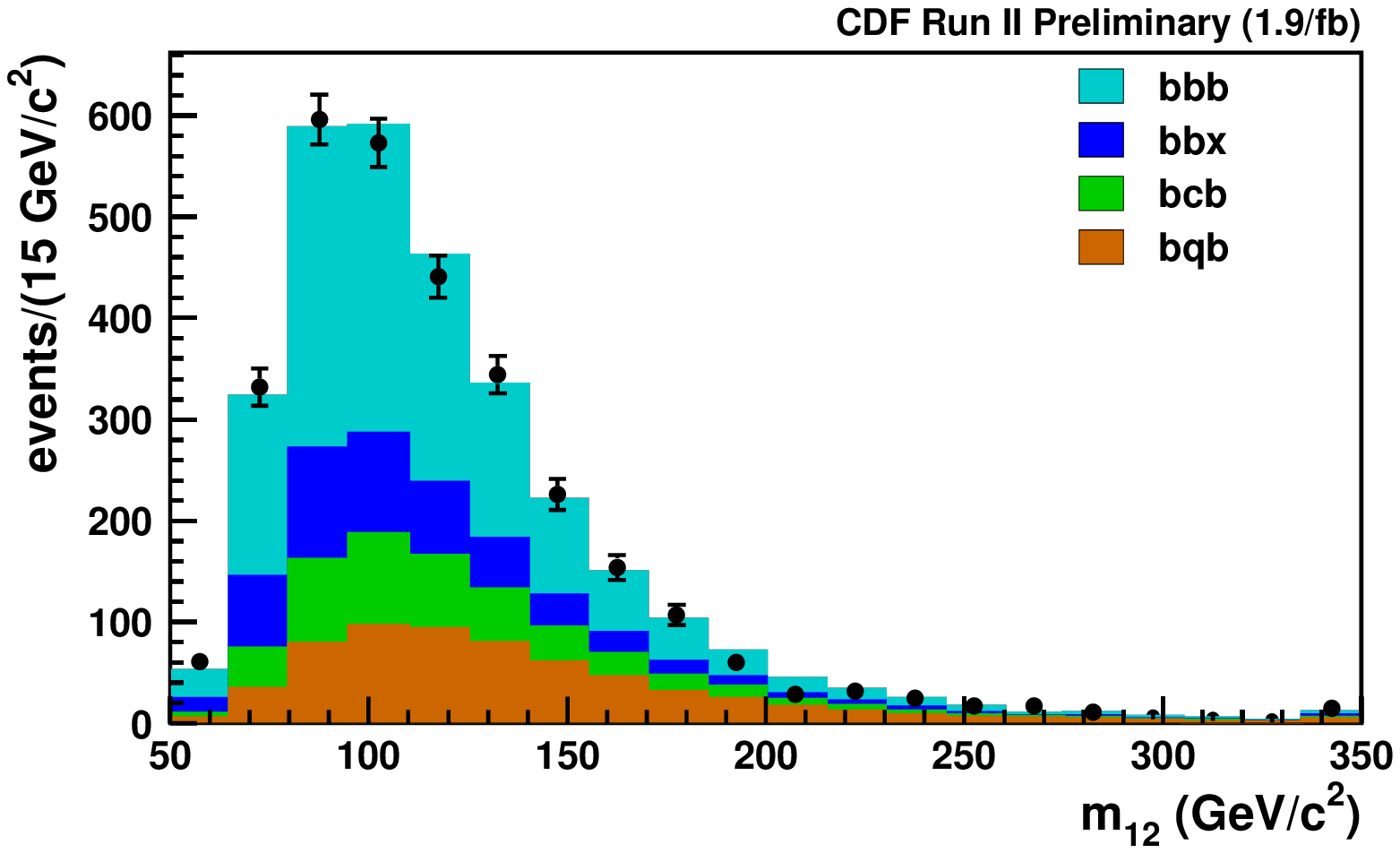} \hspace*{0.5cm}
            \includegraphics[width=0.40\textwidth]{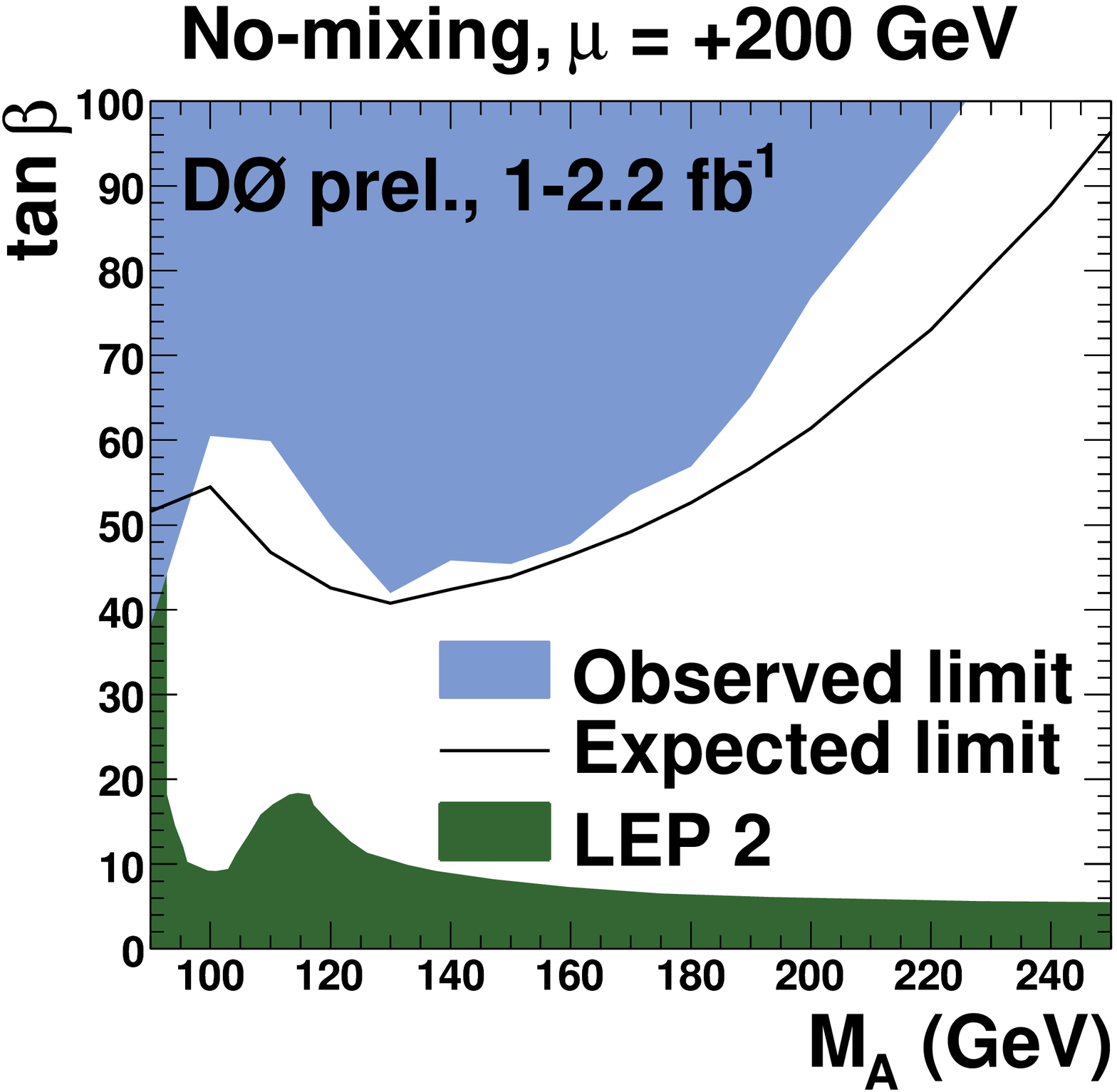}}
\caption{Left: the mass distribution of two $b-$tagged jets in events with three $b-$tagged jets from CDF. Right: 95\% C.L. limit on the $\tan\beta$ vs $m_\phi$ ($M_A$ in the figure) plane from the D\O\ $\phi\to\tau\tau$ analysis.}
\label{fig:mssm}
\end{figure}

For the same reason as $H\to b\bar{b}$ in the standard model case, inclusive search of $\phi\to b\bar{b}$ is impractical due to the large QCD $b\bar{b}$ background. The extra $b-$jet makes $b\phi\to b\bar{b}b$ final state manageable. Since there is neither a high $p_T$ lepton nor a significant $\met$, signal events have to be triggered by multijet triggers. QCD multijet production with $b$ quarks constitutes the majority of the background and its contribution is determined from data. The analysis selects events with three $b-$tagged jets and searches for mass bumps of two $b-$tagged jets. Therefore good modeling of the dijet mass distribution is critical. As an example, Fig.~\ref{fig:mssm}~(left) shows the mass distribution of two $b-$tagged jets from CDF demonstrating a good agreement between the distribution of the signal sample and the sum of distributions from various background sources. Neither CDF nor D\O\ has seen any significant excess beyond the expected backgrounds in datasets of 1.8 and 2.6~fb$^{-1}$ respectively. A significant region of the supersymmetry parameter space with a large $\tan\beta$ value is excluded.

$\phi\to\tau\tau$ is another important search final state at the Tevatron. With at least one $\tau$ decaying to an electron or a muon, the final state is sufficiently clean that the signal can be identified over backgrounds, comprised mostly of $Z\to\tau\tau$, $W+$jets and multijets. Efficient and clean identification of hadronically decaying $\tau$'s is the key for the analysis. Both CDF and D\O\ have developed sophisticated $\tau$ algorithms based on large samples of $W\to\tau\nu$ and $Z\to\tau\tau$ events. The $\phi\to\tau\tau$ analysis searches for excesses in the visible mass distribution of the two $\tau$'s. D\O\ has analyzed 2.2~fb$^{-1}$ of the data and sees no evidence for $\phi\to\tau\tau$ production. The limit in one of the MSSM scenarios from this analysis is shown in Fig.~\ref{fig:mssm}~(right). Neither has any signal been seen by CDF from a dataset of 2.0~fb$^{-1}$. D\O\ has also searched for the $\phi$ production in $b\phi\to b\tau\tau$ final state. The extra $b-$jet provides an additional tool for background rejection. This analysis yields a similar sensitivity as the inclusive $\phi\to\tau\tau$ analysis.

The combination of the CDF and D\O\ results is ongoing. We expect significant improvement in the limit once it is done.

\section{Summary and Prospect}
The quest for Higgs bosons in both standard and non-standard models is at the core of the Fermilab Tevatron Run II physics program. Both CDF and D\O\ have a comprehensive Higgs search effort. The sensitivities continue to improve at a rate faster than the rate of data accumulation. The search has reached a tipping point, particularly for a Higgs mass around 170~GeV. For the first time, direct searches at the Tevatron are contributing to our knowledge of the Higgs boson mass beyond those from LEP. 

The Fermilab Tevatron collider continues to shatter luminosity records. It is scheduled to run in 2009. A decision, likely a positive one, on 2010 running is expected next spring. At the current pace, it is very likely that both experiments will have over 8~fb$^{-1}$ of data recorded by the end of 2010. These huge datasets along with ever-improving analyses will undoubtedly lead to further improvements in sensitivities for the Higgs search. It appears that the Tevatron will be able to probe the standard model Higgs in a mass region between $145-185$~GeV. The low mass region is as expected more challenging, just as it will be at the LHC. Nevertheless, with potential improvements the Tevatron could well have sensivities in this region as well. For supersymmetric Higgs bosons, the Tevatron will continue to explore the parameter space which may lead to either evidence/discovery or large exclusion regions.  

I conclude by saying that the best Tevatron Run II results are yet to come. I thank the CDF and D\O\ Collaborations for the opportunity to present these exciting results and the Fermilab Tevatron staff for delivering outstanding integrated luminosity! I wish to express my gratitude to G.~Davies, M.~Herndon, and D.~Lincoln for commenting on this proceeding.

\end{document}